\documentclass{iau307}
\usepackage{graphicx}
\usepackage{natbib}
\usepackage{caption}
\bibpunct{(}{)}{;}{a}{}{,}

\title[The magnetic field of $\zeta$\,Ori\,A] %% give here short title %%
{The magnetic field of $\zeta$\,Ori\,A}

\author[Blaz\`ere, Neiner et al.]   %% give here short author list %%
{A. Blaz\`ere$^{1}$, C. Neiner$^1$, J-C. Bouret$^2$, A. Tkachenko$^3$\footnote{Postdoctoral Fellow of the Fund for Scientific Research (FWO), Flanders, Belgium}
 \and the MiMeS collaboration}

\affiliation{$^1$LESIA, Observatoire de Paris, UMR 8109 du CNRS, UPMC, Universit\'e Paris-Diderot, 5 place Jules Janssen, 92195 Meudon, France \\ email: {\tt aurore.blazere@obspm.fr} \\[\affilskip]
$^2$ Aix-Marseille University, CNRS, LAM, UMR 7326, 13388 Marseille, France\\
$^3$Instituut voor Sterrenkunde, KU Leuven, Celestijnenlaan 200D, B-3001 Leuven, Belgium}

\pubyear{2014}
\volume{307} 
\pagerange{}
\jname{New windows on massive stars: asteroseismology, interferometry, and spectropolarimetry}
\editors{G. Meynet, C. Georgy, J.H. Groh \& Ph. Stee, eds.}
\begin{document}

\maketitle

\begin{abstract}
Magnetic fields play a significant role in the evolution of massive stars. About
7\% of massive stars are found to be magnetic at a level detectable with current
instrumentation \citep{wade13} and only a few magnetic O stars are known. Detecting magnetic field in O stars is
particularly challenging because they only have few, often broad, lines to
measure the field, which leads to a deficit in the knowledge of the basic
magnetic properties of O stars. We present new spectropolarimetric Narval
observations of $\zeta$\,Ori\,A. We also provide a new analysis of both the new
and older data taking binarity into account. The aim of this study was to
confirm the presence of a magnetic field in $\zeta$\,Ori\,A. We identify that
it belongs to $\zeta$ Ori Aa and  characterize it.
\end{abstract}

\firstsection % if your document starts with a section,
\section{Introduction}

A magnetic field seems to have been detected in the supergiant O9.5I star
$\zeta$\,Ori\,A \citep{bouret08}. This magnetic field is the weakest
ever reported in a massive star. Thanks to this measurement of the magnetic
field one can locate $\zeta$\,Ori\,A in the magnetic confinement-rotation
diagram \citep{Petit13}. $\zeta$\,Ori\,A is the only known magnetic massive star
with a confinement parameter below 1, i.e. without a magnetosphere. However,
\cite{hummel13} recently found that $\zeta$\,Ori\,A is a O9.5I+B1IV binary star
with a period of 2687.3 $\pm$ 7.0 days, so the magnetic field of $\zeta$\,Ori\,A and its confinement parameter
might have been wrongly estimated. 
 
\section{Spectropolarimetric analysis}

\begin{figure}[!t]
\begin{center}
\includegraphics[width=6.5cm,clip]{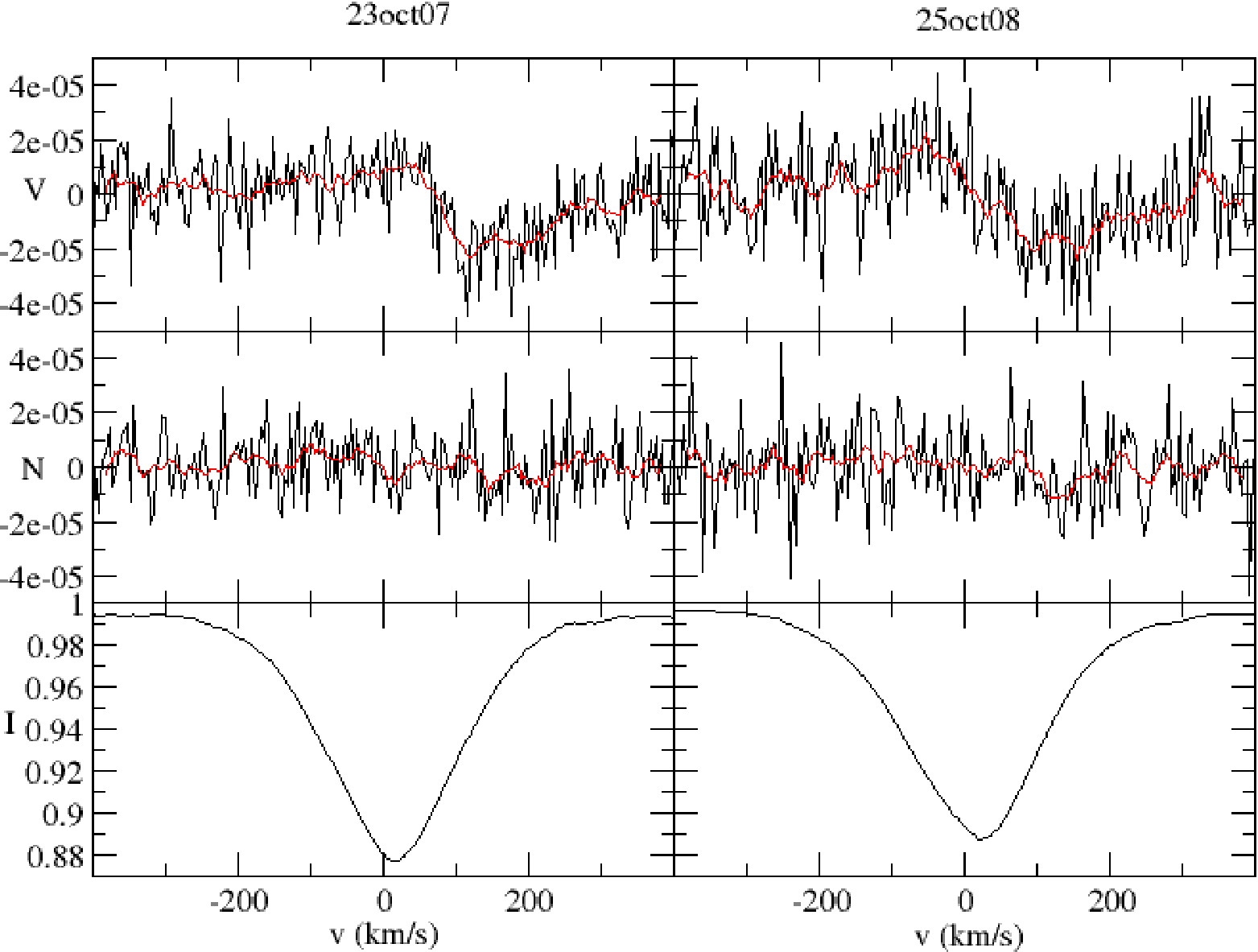}
\includegraphics[width=6.6cm,clip]{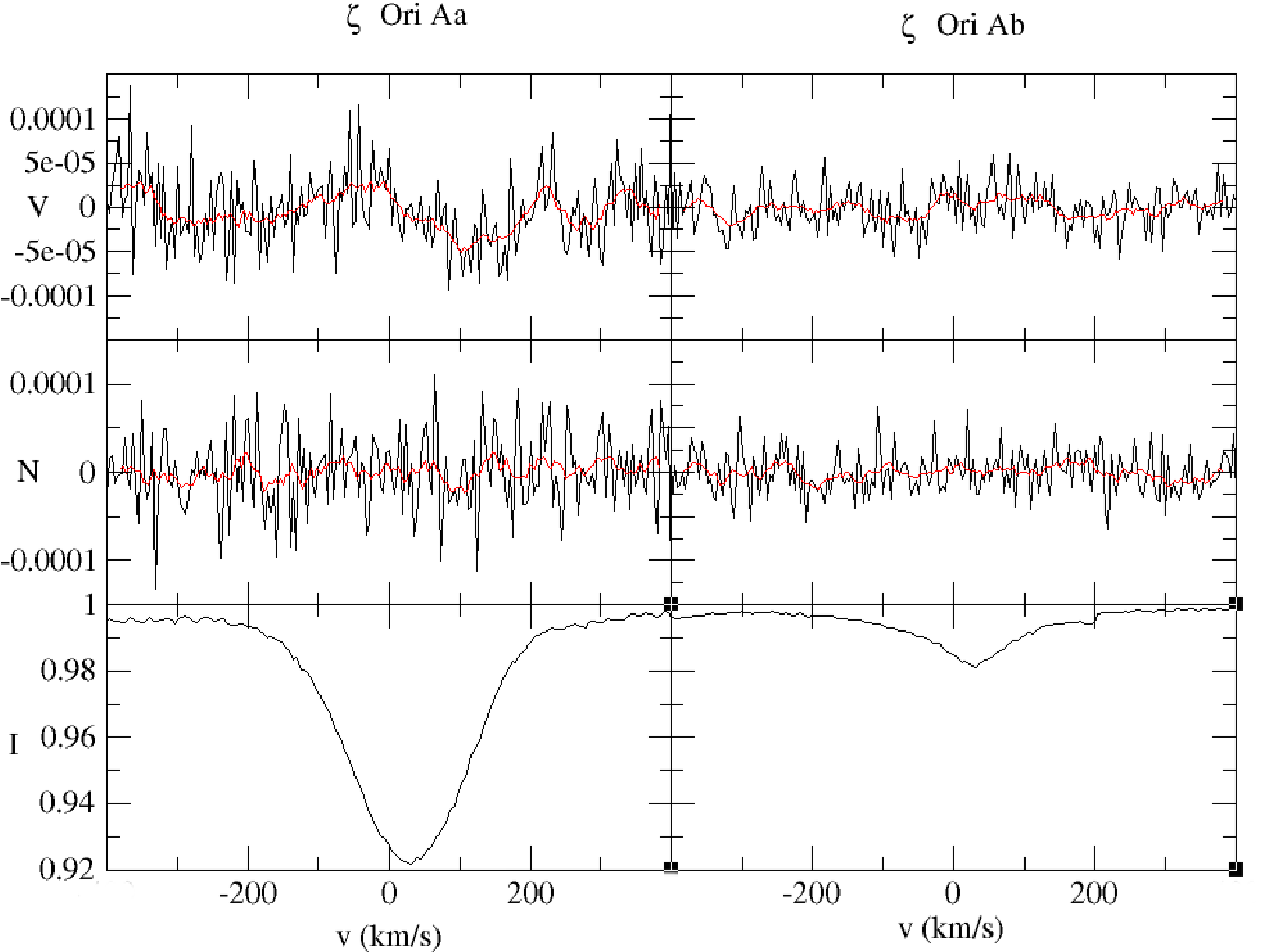}
\caption{Left: Examples of I (bottom), null N (center) and Stokes V (top) profiles for two nights. Right: I (bottom), null N (center) and Stokes V (top) profiles for October 25, 2008, for $\zeta$\,Ori\,Aa (left) and $\zeta$\,Ori\,Ab (right). The red lines represent the averaged signal.}
\label{lsd}
\end{center}
\end{figure}

Spectropolarimetric data were obtained with Narval at TBL between 2008 and 2012.
To improve the signal-to-noise ratio, we have coadded spectra obtained on the
same night, leading to 36 average spectra. We applied the well-known and commonly used Least-Squares Deconvolution (LSD) technique \citep{donati97} on each average spectrum. The mask is created from a list of lines extracted from VALD \citep{piskunov95, kupka99} using the effective temperature $T_{\rm eff}$=30000K and the log g=3.25. We then
extracted LSD Stokes I and V profiles for each night as well as null (N)
polarization profiles to check for spurious signatures. We see clear Zeeman
signatures some nights (see left panel of Fig.~\ref{lsd}). $\zeta$\,Ori\,A is thus confirmed to be magnetic. However, $\zeta$\,Ori\,A is a binary and we ignore which component of the binary is magnetic, or if both components are magnetic. To provide an answer to this question we must disentangle the composite spectra. 

\section{Disentangling the spectra}

We used the Fourier-based formulation of the spectral disentangling method
\citep{hadrava95} as implemented in the {\sc FDBinary} code \citep{ilijic04} to
try to separate the individual spectra of both components of the
$\zeta$\,Ori\,A binary system. This method failed due to a bad phase coverage of the binary period.

Therefore we used a second method to disentangle the spectra: for each component
of $\zeta$\,Ori\,A we computed a synthetic spectrum with their respective
temperature and log g. Based on these synthetic spectra, we checked which lines
in the observations comes from only one of the component. We created two distinct masks
with lines from only one of the component and we used LSD with these masks. We
find that the magnetic field is present in $\zeta$\,Ori\,Aa and not in $\zeta$\,Ori\,Ab (see right panel of
Fig.~\ref{lsd}).

\section{Conclusion}

We confirm that $\zeta$\,Ori\,A is a magnetic star. The magnetic field is
present in the supergiant $\zeta$\,Ori\,Aa and no magnetic field is detected in its companion
$\zeta$\,Ori\,Ab.

\bibliographystyle{iau307}
\bibliography{biblio.bib}

\begin{thebibliography}{}

\bibitem[\protect\astroncite{{Bouret} et~al.}{2008}]{bouret08}
{Bouret}, J.-C., {Donati}, J.-F., {Martins}, F., {et~al.} 2008,
\newblock {\em \mnras} 389, 75

\bibitem[\protect\astroncite{{Donati} et~al.}{1997}]{donati97}
{Donati}, J.-F., {Semel}, M., {Carter}, B.~D., {Rees}, D.~E., \& {Collier
  Cameron}, A. 1997,
\newblock {\em \mnras} 291, 658

\bibitem[\protect\astroncite{{Hadrava}}{1995}]{hadrava95}
{Hadrava}, P. 1995,
\newblock {\em \aaps} 114, 393

\bibitem[\protect\astroncite{{Hummel} et~al.}{2013}]{hummel13}
{Hummel}, C.~A., {Rivinius}, T., {Nieva}, M.-F., {et~al.} 2013,
\newblock {\em \aap} 554, A52

\bibitem[\protect\astroncite{{Ilijic} et~al.}{2004}]{ilijic04}
{Ilijic}, S., {Hensberge}, H., {Pavlovski}, K., \& {Freyhammer}, L.~M. 2004,
\newblock Vol. 318 of {\em ASPCS}, p. 111

\bibitem[\protect\astroncite{{Kupka} \& {Ryabchikova}}{1999}]{kupka99}
{Kupka}, F. \& {Ryabchikova}, T.~A. 1999,
\newblock {\em Publications de l'Observatoire Astronomique de Beograd} 65, 223

\bibitem[\protect\astroncite{{Petit} et~al.}{2013}]{Petit13}
{Petit}, V., {Owocki}, S.~P., {Wade}, G.~A., {et~al.} 2013,
\newblock {\em \mnras} 429, 398

\bibitem[\protect\astroncite{{Piskunov} et~al.}{1995}]{piskunov95}
{Piskunov}, N.~E., {Kupka}, F., {Ryabchikova}, T.~A., {Weiss}, W.~W., \&
  {Jeffery}, C.~S. 1995,
\newblock {\em \aaps} 112, 525

\bibitem[\protect\astroncite{{Wade} et~al.}{2013}]{wade13}
{Wade}, G.~A., {Grunhut}, J., {Alecian}, E., {et~al.} 2013,
\newblock {\em IAUS 302, ArXiv e-prints 1310.3965}

\end{thebibliography}

\end{document}